\documentclass[letterpaper, 10 pt, conference]{ieeeconf}  

\IEEEoverridecommandlockouts                              
\usepackage{amsmath} 
\usepackage{amssymb}  
\usepackage{color}
\usepackage{multirow}
\usepackage{array}
\usepackage{algorithm}
\usepackage{algpseudocode}
\usepackage{mathrsfs,comment}
\usepackage{graphicx}
\usepackage{cite}

\makeatletter
\let\NAT@parse\undefined
\makeatother

\usepackage[
    colorlinks=true,
    linkcolor=blue,
    citecolor=blue,
    urlcolor=blue
]{hyperref}

\title{\LARGE \bf
An extended Perron--Frobenius operator filter \\ for nonlinear state estimation
}

\author{Yuta Miwa, Yoshihiko Susuki, and Shunji Kotsuki
\thanks{Y.~Miwa and Y.~Susuki are with the Department of Electrical, Electronic, and Digital Science and Engineering, Kyoto University
         Katsura, Nishikyo-ku, Kyoto, 615-8510, Japan 
         (e-mail: \{y-miwa@dove.kuee., susuki.yoshihiko.5c@\}kyoto-u.ac.jp).}
\thanks{S.~Kotsuki is with the Institute for Advanced Academic Research, 
         Chiba University, Yayoi-cho 1-33, Inage-ku, Chiba, 263-8522, Japan
         (e-mail: shunji.kotsuki@chiba-u.jp).}    
}

\begin{document}

\maketitle
\thispagestyle{empty}
\pagestyle{empty}

\begin{abstract}
We propose an extended Perron--Frobenius Operator Filter (ePFOF) for nonlinear state estimation. 
The method learns the Perron--Frobenius operator, an infinite-dimensional linear operator fully preserving properties of a nonlinear dynamical system, using the extended Dynamic Mode Decomposition (eDMD). 
This enables us to explicitly account for non-Gaussian distributions exhibited by the nonlinear system within a linear-operator representation, while retaining the freedom to choose basis functions in eDMD.
Through two numerical examples, we show that the extended PFOF achieves high computational efficiency 
and high estimation accuracy by exploiting the flexibility in the choice of basis functions in eDMD. 
\end{abstract}


%
\section{Introduction}

\begingroup
\renewcommand{\thefootnote}{}
\footnotetext{This is a revised version of the manuscript submitted to Nonlinear Theory and Its Applications, IEICE (NOLTA).}
\endgroup

The so-called state estimation is a fundamental technique in nonlinear engineering 
to infer states of nonlinear dynamical systems  
from limited available information, called observations. 
When the states and observations are modeled as random variables, the estimation problem is generally formulated 
within the Bayesian filtering framework~\cite{sarkka2023bayesian,jazwinski1970}. 
For nonlinear and non-Gaussian systems, 
various Bayesian filters, 
such as the extended Kalman Filter (EKF)~\cite{jazwinski1970}, 
Ensemble Kalman Filter (EnKF)~\cite{evensen2003}, 
and Particle Filter (PF)~\cite{gordon1993}, have been widely used. 
While these methods can be effective 
for systems with weak nonlinearity or low dimensionality, 
extending them to high-dimensional systems 
with strong nonlinearity and non-Gaussian characteristics 
is challenging~\cite{bocquet2010beyond,snyder2008obstacles}.

The linear-operator framework for nonlinear dynamical systems~\cite{mauroy2020,brunton2022} 
has recently attracted significant interest as a new direction 
in nonlinear dynamical analysis~\cite{mezic2013,susuki2016applied}, control~\cite{otto2021,bevanda2021}, and state estimation~\cite{surana2016linear,netto2018robust,dutta2013nonlinear,liu2024perron}. 
In this framework, the time evolution of states of a nonlinear system is represented through the time evolution of scalar-valued functions defined on the state space. 
In particular, when the scalar-valued function is chosen as a probability density function, its time evolution 
is governed by the Perron--Frobenius Operator (PFO)~\cite{lasota2013chaos,mauroy2020}.
The PFO acts linearly on the space of probability density functions (i.e., an infinite-dimensional function space), 
thereby enabling probabilistic analysis and prediction of the original nonlinear system within a linear-operator representation. 

PFO-based Bayesian filtering has been studied 
for nonlinear and non-Gaussian state estimation.
In this approach, the PFO is used to propagate 
the states' probability density functions during the prediction 
step of Bayesian filtering.
To the best of the authors' knowledge,
Bayesian filtering using the PFO 
was first proposed by Dutta et al.~\cite{dutta2013nonlinear}.
Subsequently, Liu and Jiang~\cite{liu2024perron}
proposed a framework for PFO-based Bayesian filtering,
including the estimation of a finite-dimensional matrix representation of the PFO,
which is referred to as the Perron--Frobenius Operator Filter (PFOF).

Numerical methods for finite-dimensional approximation of the PFO
include Ulam's method~\cite{ulam1960collection,klus2016numerical,junge2009discretization}
and the extended Dynamic Mode Decomposition (eDMD)~\cite{klus2016numerical,goswami2018constrained,williams2015data}.
In Ulam's method,
the PFO is approximated
on a finite-dimensional subspace spanned by indicator functions defined on a grid on the state space.
The PFOF proposed in~\cite{liu2024perron}
employs Ulam's method to construct a finite-dimensional approximation of the PFO.
The eDMD was originally developed to approximate the Koopman operator (that is, an adjoint operator of the PFO) in \cite{williams2015data} and  
can approximate the PFO by exploiting their adjoint relationship as shown in \cite{klus2016numerical}. 
While Ulam's method employs indicator functions for basis functions,
eDMD allows free choice of basis functions.
This flexibility enables the design of finite-dimensional approximation spaces
according to the dynamical properties of the target system.

Recently, the DATO framework,
which combines a PFO estimated using kernel eDMD with Bayesian filtering,
has been described in~\cite{donno2026classical}.
In DATO, a spectral representation of the PFO is estimated
from training samples using kernel eDMD,
and the Bayesian updates are performed
in the finite-dimensional function space induced by the training data.

In contrast, we investigate an approach
in which the PFO is approximated using eDMD
in a finite-dimensional function space spanned by explicitly prescribed
basis functions, 
and Bayesian updates 
are performed in this finite-dimensional space. This formulation enables 
the explicit design of basis functions for which we can consider properties of the underlying system.

This paper extends our preliminary work presented in~\cite{miwa2026noltasociety} 
by providing a detailed formulation and comprehensive numerical evaluation of the proposed method. 
Specifically, we present the computational algorithm of ePFOF, including the empirical Galerkin implementation of the Bayesian analysis step, 
and provide quantitative evaluations of the estimation accuracy and computational cost of the proposed and benchmark methods.
We also introduce a numerical example with basis functions synthesized to reflect the dynamical properties of the target system, demonstrating that the flexibility of eDMD in choosing basis functions can be leveraged to improve estimation accuracy.

The rest of this paper is organized as follows. 
Section~\ref{sec:preliminaries} provides the preliminaries. 
Section~\ref{sec:ePFOF} formulates the ePFOF and presents its computational algorithm. 
Section~\ref{sec:performance_eval} evaluates the performance of ePFOF on two strongly nonlinear systems, 
and Section~\ref{sec:conclusion} concludes this paper.

\section{Preliminaries}
\label{sec:preliminaries}

\subsection{Perron--Frobenius Operator}
Let $(X, \mathcal{B}, \mu)$ be a measure space, 
where $X \subseteq \mathbb{R}^n$ is a state space, 
$\mathcal{B}$ is a $\sigma$-algebra of $X$,  
and $\mu$ is a reference measure. 
The standard $\mathcal{L}^2$ space $ \mathcal{L}^2(X,\mathcal{B},\mu)=:\mathcal{F}$ is introduced 
as the space of measurable functions $f:X\to\mathbb{R}$ 
such that $\int_X |f|^2 \mathrm{d}\mu < \infty$.
Throughout this paper, we consider a discrete-time dynamical system given by
\begin{align}
   x_{k+1} &= T(x_k),
   \label{eq:nonlinear_system}
\end{align}
where $k \in \mathbb{Z}$, $x_k\in X$, and $T:X\to X$ is a measurable, nonsingular nonlinear map. 
The PFO
$\mathscr{P}:\mathcal{F}\to\mathcal{F}$ is defined for the nonlinear system \eqref{eq:nonlinear_system} by
\begin{align*}
   \int_{T^{-1}(A)} f(x)\mathrm{d}\mu = \int_A (\mathscr{P}f)(x)\mathrm{d}\mu, \quad \forall A\in\mathcal{B}.
\end{align*}
This PFO is an infinite-dimensional but linear operator fully preserving properties of the nonlinear system \eqref{eq:nonlinear_system} (see \cite{lasota2013chaos,mauroy2020}). 
If $x_k$ is modeled as a random variable on $X$ that
admits a Probability Density Function (PDF) $p_{x_k}$ with respect to $\mu$, 
then the PFO governs its temporal evolution as
\begin{align}
   p_{x_{k+1}} = \mathscr{P} p_{x_k}.
   \label{eqn:PDF_evolution}
\end{align}

\subsection{Extended Dynamic Mode Decomposition 
\label{subsubsec:eDMD}}
The eDMD in \cite{klus2016numerical} is a numerical scheme of estimating a finite-dimensional approximation of the PFO. 
Let $\{\psi_1,\dots,\psi_N\}$ be a set of $N$ basis functions spanning 
a finite-dimensional subspace of $\mathcal{F}$ as $\mathcal{V} := \text{span}\{\psi_1,\dots,\psi_N\}$. 
Then, the finite-dimensional approximation of the PFO,
$\mathscr{P}_\mathcal{V}: \mathcal{V}\to\mathcal{V}$,  
is represented by a finite-dimensional matrix $P$ satisfying
\begin{align*}
   \begin{bmatrix}
      \mathscr{P}_\mathcal{V} \psi_1, 
      \ldots, 
      \mathscr{P}_\mathcal{V} \psi_N
   \end{bmatrix}
   =
   \begin{bmatrix}
      \psi_1, 
      \ldots, 
      \psi_N
   \end{bmatrix}P.
\end{align*}
Given $M$ samples of the state, $x^{(1)}, \ldots, x^{(M)} \in X$, 
and their one-step images $x^{+(l)} = T(x^{(l)})$ ($l=1, \ldots, M$), 
let us define the Gram matrix $G$ 
and shifted covariance matrix $G_\mathrm{SC}$ as follows:
\begin{align*}
   G := \frac{1}{M}\Psi_X \Psi_X^\top,\quad G_\mathrm{SC} := \frac{1}{M}\Psi_X \Psi_{X}^{+\top},
\end{align*}
with
\begin{align*}
  &\Psi_X := 
  \begin{bmatrix}
    \psi_1(x^{(1)})  &\cdots &\psi_1(x^{(M)})\\ 
    \vdots &\ddots &\vdots\\ 
    \psi_N(x^{(1)})  &\cdots &\psi_N(x^{(M)})\\ 
  \end{bmatrix},\\ 
  &\Psi_X^+ := 
  \begin{bmatrix}
    \psi_1(x^{+(1)}) &\cdots &\psi_1(x^{+(M)})\\ 
    \vdots &\ddots &\vdots\\ 
    \psi_N(x^{+(1)}) &\cdots &\psi_N(x^{+(M)})\\ 
  \end{bmatrix}.
\end{align*}
The symbol $\top$ above stands for the transpose operation of matrices. 
Then, the PFO matrix $P$ can be approximated from the data as
$
   P \approx (G + \lambda I_{N})^{-1} G_\mathrm{SC}^\top,
$
where $I_N$ is the $N \times N$ identity matrix, and $\lambda > 0$ is the Tikhonov regularization parameter.
In the numerical examples below, we fix $\lambda = 10^{-8}$.

\subsection{Bayesian Filter}
Consider an output (observation) equation for the nonlinear system \eqref{eq:nonlinear_system} as
\begin{align}
   y_k = h(x_k) + v_k, \label{eq:y}
\end{align}
where $y_k \in \mathbb{R}^m$ is an $m$-dimensional observation and $v_k \in \mathbb{R}^m$ is an observation noise.
The function $h : X \to \mathbb{R}^m$ represents a nonlinear model of the observation.

Using a probabilistic setting, let us denote a collection of observations from the initial time $0$ to $k$ 
by $\mathbf{y}^k := \{y_0^o,\ldots,y_k^o\}$,
where $y^o$ stands for a realization of a random variable $y$.
Let $\rho_{k|k-1}$ and $\rho_{k|k}$ be the conditional PDFs of $x_k$ given 
$\mathbf{y}^{k-1}$ (\emph{prior} distribution) and $\mathbf{y}^{k}$ (\emph{posterior} distribution), respectively.
In addition, let $\ell_k(x)$ be the likelihood function corresponding to the observation $y_k^o$ at time $k$.

According to \eqref{eqn:PDF_evolution}, the update of Bayesian filtering consists of the following two steps \cite{liu2024perron}:
\par\vspace{0.5\baselineskip}
\noindent\textbf{Prediction:}
\vspace{-0.5\baselineskip}
\begin{align}
   \rho_{k|k-1}(x)
     &= \int_X \delta\!\left(x - T(\chi)\right)
        \rho_{k-1|k-1}(\chi)\, \mathrm{d}\chi \nonumber \\ 
     &=: (\mathscr{P}\rho_{k-1|k-1})(x),
        \label{eq:bayes_prediction}
\end{align}
\vspace{-1.75\baselineskip}
\par\noindent\textbf{Analysis:}\vspace{-0.5\baselineskip}
\begin{align}
   \rho_{k|k}(x)
     &= \frac{\ell_k(x)\rho_{k|k-1}(x)}
        {\int_X \ell_k(\xi)\rho_{k|k-1}(\xi)\,\mathrm{d}\xi} \nonumber \\ 
     &=: (\mathscr{L}_k\rho_{k|k-1})(x).
        \label{eq:bayes_analysis}
\end{align}
Here, $\mathscr{L}_k$ is called the likelihood operator~\cite{liu2024perron}, which maps the prior distribution $\rho_{k|k-1}$ to the posterior one $\rho_{k|k}$, 
and $\delta(x)$ is the Dirac delta function supported at $0\in X$.
Consequently, the Bayesian filter can be written as the following operator's  composition:
\begin{align}
   \rho_{k|k} = \mathscr{L}_k \mathscr{P}\rho_{k-1|k-1}. \label{eq:bayes-functional}
\end{align}

\section{Extended Perron--Frobenius Operator Filter}
\label{sec:ePFOF}

\subsection{The Original Formulation in \cite{liu2024perron}}
The Perron--Frobenius Operator Filter (PFOF) is a finite-dimensional approximation of the Bayesian filter \eqref{eq:bayes-functional} using the PFO, given as
\begin{align}
   \rho^{\mathcal{U}}_{k|k} = \mathscr{L}^\mathcal{U}_k \mathscr{P}_\mathcal{U} \rho^{\mathcal{U}}_{k-1|k-1},
   \; \rho^{\mathcal{U}}_{0|0} = \mathscr{L}^\mathcal{U}_0\rho_{0}, \; \rho^\mathcal{U}_{k|k}\in\mathcal{U},
   \label{eq:pfof}
\end{align}
where $\mathscr{P}_\mathcal{U}$ is a finite-dimensional approximation of the PFO $\mathscr{P}$ 
obtained via Ulam's method~\cite{klus2016numerical}, 
$\mathcal{U}$ is a function space spanned by indicator functions defined on a grid on $X$,
$\mathscr{L}_k^\mathcal{U}$ is the finite-dimensional likelihood operator
that updates the coefficients of the indicator functions based on the likelihood values, as described in \cite{liu2024perron} and $\rho_0$ is an initial distribution.

\subsection{Main Idea}
\label{sec:PFOF}
We newly propose an extended Perron--Frobenius Operator Filter (ePFOF) that exploits the  eDMD~\cite{klus2016numerical} to approximate the PFO $\mathscr{P}$ in the space $\mathcal{V}$ based on freely chosen basis functions:
\begin{align}
   \rho^{\mathcal{V}}_{k|k} = \mathscr{L}^\mathcal{V}_k \mathscr{P}_\mathcal{V} \rho^{\mathcal{V}}_{k-1|k-1},
   \; \rho^{\mathcal{V}}_{0|0} = \mathscr{L}^\mathcal{V}_0\rho_{0}, \; \rho^\mathcal{V}_{k|k}\in\mathcal{V},
   \label{eq:epfof}
\end{align}
where the finite-dimensional approximation $\mathscr{P}_\mathcal{V}$ is obtained via eDMD, 
$\mathscr{L}_k^\mathcal{V}$ is the finite-dimensional approximation of the likelihood operator, 
which is implemented through an empirical Galerkin projection onto $\mathcal{V}$, followed by normalization.

It should be noted that, although the exact Perron--Frobenius operator 
preserves the non-negativity of probability density functions, 
its finite-dimensional approximation obtained by eDMD does not generally 
guarantee this non-negativity. 
Therefore, $\rho^\mathcal{V}_{k|k}$ and $\rho^\mathcal{V}_{k|k-1}$ in (8) 
should be interpreted as finite-dimensional approximations 
of the posterior and prior densities, 
rather than probability density functions in the strict sense.

The update procedure of ePFOF is summarized below. 
We define the basis vector $\Psi(x) := [\psi_1(x),\ldots,\psi_N(x)]^\top$, 
where $(\cdot)^\top$ stands for the transpose operation of vectors.
Since $\mathscr{P}_\mathcal{V}$ acts on the finite-dimensional space $\mathcal{V}$ with this basis vector, 
the posterior distribution $\rho^\mathcal{V}_{k|k}(x)$ and the prior distribution $\rho^\mathcal{V}_{k|k-1}(x)$ can be represented as
\begin{align*}
   \rho^\mathcal{V}_{k|k}(x) = c_{k|k}^\top \Psi(x), \quad
   \rho^\mathcal{V}_{k|k-1}(x) = c_{k|k-1}^\top \Psi(x),
\end{align*}
where $c_{k|k}, c_{k|k-1}\in\mathbb{R}^N$ are the corresponding coefficient vectors for the $\Psi(x)$-basis. 
The filtering update is implemented on the two coefficient vectors.

\subsubsection*{Prediction:}
By using the PFO matrix $P$ for 
$\mathscr{P}_{\mathcal{V}}$, the prediction step \eqref{eq:bayes_prediction} is simply performed as
\begin{align}
   c_{k|k-1} = P\,c_{k-1|k-1}.
\end{align}

\subsubsection*{Analysis:} 
The analysis step \eqref{eq:bayes_analysis} is performed with the Galerkin projection. 
For this, it is recalled here that the likelihood function $\ell_k(x)$ 
is defined through \eqref{eq:y} with the observation model and the noise distribution; 
therefore, $\ell_k(x)$ does not generally belong to the finite-dimensional subspace $\mathcal{V}$. 
Consequently, due to the multiplication by $\ell_k$, the likelihood operator $\mathscr{L}_k$ in \eqref{eq:bayes_analysis}
does not map $\mathcal{V}$ into itself.
To perform our update inside $\mathcal{V}$, the unnormalized posterior, defined as 
\begin{align}
   \rho_k^\ast(x) := \ell_k(x)\rho^\mathcal{V}_{k|k-1}(x)
   = \ell_k(x)\,c_{k|k-1}^\top \Psi(x),  \label{eq:rho-star}
\end{align}
is projected onto $\mathcal{V}$ 
via an empirical Galerkin projection 
using the 
same samples of the state
$x^{(1)},\ldots,x^{(M)}$ on $X$
as those used to construct the eDMD approximation. The sampling distributions used to generate these training samples are specified for each numerical example in Section \ref{sec:performance_eval}.
As a result, the coefficient vector $c^*_k$ minimizing the $\mathcal{L}^2$ norm  $\|\rho_k^\ast(x) - c^{\ast\top}_k\Psi(x)\|_{\mathcal{L}_2}$ is obtained by solving the linear equation
\begin{align}
   LL^\top c^*_k  = \frac{1}{M}\Psi_X \left({e}_k \odot \Psi_X^\top c_{k|k-1}\right), \label{eq:c_ast}
\end{align}
where $\odot$ stands for the Hadamard product, $L$ is the Cholesky factor of $G$, and
\begin{align}
   {e}_k := 
   \begin{bmatrix}
     \ell_k(x^{(1)}),\ldots,\ell_k(x^{(M)})
   \end{bmatrix}^\top.
   \label{eq:p_k}
\end{align}
The detailed derivation of \eqref{eq:c_ast} is provided in the Appendix. 
Finally, the posterior distribution is obtained by normalization as
\begin{align*}
   \rho^\mathcal{V}_{k|k}(x) = c_{k|k}^\top \Psi(x),\quad c_{k|k}
   :=\frac{c_k^*}{c_k^{*\top}{b}},
\end{align*}
with ${b} := [b_1\ldots b_N]^\top$ and $b_i:=\int_{X}\psi_i(x)\mathrm{d}x$. 
Since the basis functions are predefined, these integrals can be numerically pre-computed offline.

The initial coefficient vector $c_{0|0}$ is obtained using the 
same empirical Galerkin projection as in the analysis step. 
Specifically, the unnormalized initial posterior is given by 
$\rho^*_0(x) := \ell_0(x)\rho_0(x)$. 
Let
$r_0:=\left[\rho_0(x^{(1)}),\ldots,\rho_0(x^{(M)})\right]^\top .$
Then, the coefficient vector $c_{0}^*$ 
minimizing the $\mathcal{L}^2$ norm  $\|\rho_0^\ast(x) - c^{\ast\top}_0\Psi(x)\|_{\mathcal{L}_2}$ 
is obtained by solving
\begin{align*}
LL^\top c_{0}^*
= \frac{1}{M}\Psi_X
\left(e_0\odot r_0\right),
\end{align*}
and is normalized as
\begin{align*}
c_{0|0}=\frac{c_0^*}{c_{0}^{*\top}b}.
\end{align*}
The overall computational flow is summarized in Algorithm~\ref{alg:ePFOF}.

\begin{algorithm}[h]
  \caption{Extended Perron--Frobenius Operator Filter (ePFOF)}
  \label{alg:ePFOF}
  \begin{algorithmic}[1]
    \Statex \hspace{-\algorithmicindent} \textbf{Input (Offline):}
      \Statex PFO matrix $P$;
      Cholesky factor $L$ of $G=LL^\top$; 
      basis vector $\Psi(x)$;
      training samples $\{x^{(l)}\}_{l=1}^M$; 
      dictionary matrix $\Psi_X$;
      integral vector ${b}$; initial coefficient vector $c_{0|0}$;
   \Statex \hspace{-\algorithmicindent} \textbf{Filtering (Online):}
   \State {Initialization:} $k \leftarrow 1$.
      \State {Prediction:} $c_{k|k-1} \leftarrow P\,c_{k-1|k-1}$.
      \State {Likelihood:} define $\ell_k(x)$ for the observation 
      \Statex \hspace{\algorithmicindent} $y_k^o$ and construct the likelihood vector:
         \[
            {e}_k \leftarrow
            \begin{bmatrix}
            \ell_k(x^{(1)}),\ldots,\ell_k(x^{(M)})
            \end{bmatrix}^\top.
         \]
      \State {Analysis:} find $c^\ast_k$ by solving the linear equa- 
      \Statex \hspace{\algorithmicindent} tion
        \[
          LL^\top c_k^* = \frac{1}{M}\Psi_X \left({e}_k \odot \Psi_X^\top c_{k|k-1}\right).
        \]
      \State {Normalization:} $c_{k|k} \leftarrow \dfrac{c_k^*}{c_k^{*\top}{b}}$.
      \State $k \leftarrow k+1$
      \State Go to step 2.
  \end{algorithmic}
\end{algorithm}

\section{Performance Evaluation}
\label{sec:performance_eval}
This section evaluates the ePFOF through numerical simulations.
We consider two nonlinear systems: 
the forced Duffing equation with a chaotic attractor \cite{ueda1988forced}
and a forced swing equation with coexisting attractors 
as a model of two-machine power grid \cite{ueda1998nonlinear}.
These systems are nonlinear ordinary differential equations 
with periodic forcing terms, 
and the associated discrete-time dynamical systems (also known 
as Poincar\'e maps) 
are obtained through stroboscopic observation.
Here, we choose the Duffing equation to evaluate \emph{steady-state} 
performance of ePFOF for the two-dimensional chaotic system, 
and the swing equation to evaluate its \emph{transient-state} 
performance for the four-dimensional system,
motivated by a real-world application in power grids.

In these numerical experiments, 
the methods are quantitatively evaluated based on the root mean square error (RMSE) 
with respect to the true state 
and the online and offline computational times required for filtering. 
The RMSE at time step $k$ is defined as
$\mathrm{RMSE}_k=\sqrt{\frac{1}{n}\|\hat{x}_k-x_k\|^2}$,
where $n$ is the dimension of the state, $x_k$ is the true state at time step $k$, and $\hat{x}_k$ is the minimum mean-square error (MMSE) estimate obtained from the posterior distribution $\rho_{k|k}$.

\subsection{Forced Duffing Equation}
\begin{figure*}[tb]
\centering
\includegraphics[width=1.0\linewidth]{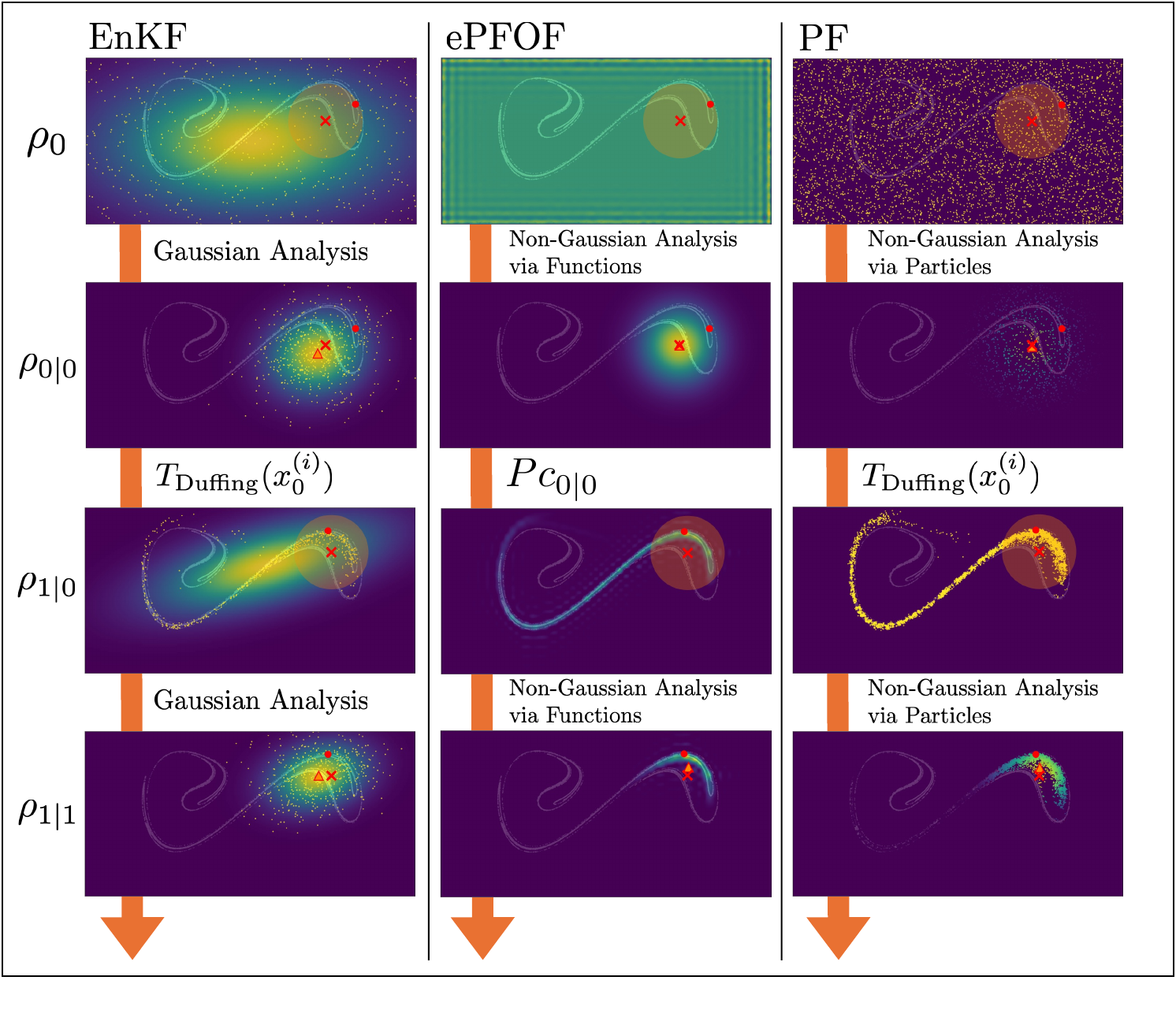}
\vspace{-20pt}
\caption{
   Update of 
   estimated distributions
   from the initial distribution to the posterior distribution 
   at $k=1$ for the forced Duffing equation: 
   from left to right, EnKF (Ensemble Kalman
   filter), ePFOF (extended Perron–Frobenius Operator Filter), and PF (Particle Filter). 
   The chaotic attractor is drawn with the semi-transparent white dots,
   and the overlaid colormap indicates the estimated distributions, 
   where the colormap is normalized to the range $[0,1]$. 
   For the EnKF and PF, colored dots
   represent the ensemble members and the particles, respectively. 
   The red semi-transparent circle in each prior distribution represents the 1.5$\sigma$ region
   of the Gaussian observation likelihood. 
   The red point, red cross, and orange triangle stand for  
   the true state, observation, and MMSE
   (Minimum Mean Square Error) estimate.
   \label{fig:dist_duffing}
}
\end{figure*}

\begin{figure}[tb]
\centering
\includegraphics[width=1.0\linewidth]{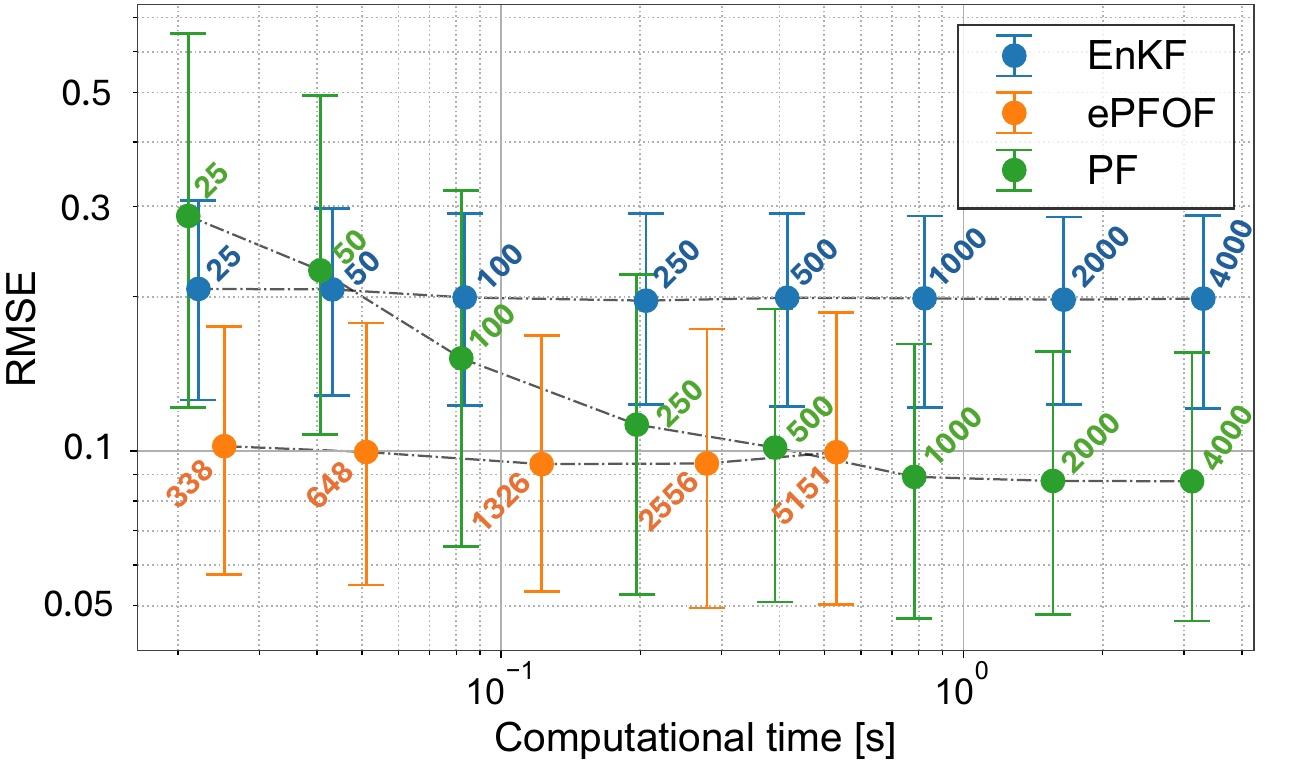}
\vspace{-10pt}
\caption{
   RMSE distributions over 200 experiments, computed using the 
   estimates from time steps $30 \leq k \leq 40$, plotted against the computational time per filtering step.
   Colored dots stand for the medians, and error bars indicate the interquartile 
   ranges (25th-75th percentiles). The numbers next to the dots 
   indicate the number of particles for PF, the number of ensemble members for EnKF, or the number of basis functions for ePFOF.
   \label{fig:rmse_duffing}
}
\end{figure}

The forced Duffing equation \cite{ueda1988forced} with a chaotic attractor, 
given as 
\begin{align}
   \dot{x} := \frac{\mathrm{d}x}{\mathrm{d}t}
   =
   \begin{bmatrix}
   x_2\\
   - 0.25 x_2 + x_1 - x_1^3 + 0.3\cos t
   \end{bmatrix}, 
   \label{eq:duffing}
\end{align}
is our target system with the two-dimensional state 
$x = [x_1, x_2]^\top$. 
The stroboscopic observation of solutions 
$\varphi(t; 0,x)$ starting from $x \in \mathbb{R}^2$ 
at time $t=0$ under the period $2\pi$ induces 
the discrete-time system 
\begin{align*}
   T_\mathrm{Duffing}: x \mapsto \varphi(2\pi;0,x), 
\end{align*}
which is a $C^\infty$-diffeomorphism. 

We suppose that the observation directly measures the states with 
Gaussian noise: 
$y_k = x_k + v_k, \quad v_k \sim \mathcal{N}_{0, 0.09I_2},$
where $\mathcal{N}_{\bar{x}, R}$ stands for the normal distribution with mean 
$\bar{x}$ and covariance matrix $R$.

\subsubsection{Set-Up}
The accuracy of eDMD strongly depends on the choice of basis functions, 
for which no systematic selection strategy has been established. 
In this example, as a simple and commonly used choice, we adopt 
Gaussian Radial Basis Functions (RBFs)
\begin{align*}
   \psi_{\bar{x},\sigma} = \exp \left(- \frac{\|x - \bar{x}\|^2}{2\sigma^2}\right), 
\end{align*}
where $\|\cdot\|$ is the Euclidean norm, 
$\bar{x}$ is the center of RBF, 
and $\sigma > 0$ controls its width. 
For the Duffing equation, we place $N$ Gaussian RBFs on a uniform grid 
over $[-2,2]\times[-1,1]$. 
To evaluate how the performance depends on $N$, 
we use several choices of grid spacing
$\Delta x  = 0.16,0.113,0.08,0.057$, and $0.04$, 
which correspond to 
$N = 338, 648, 1326, 2556$, and $5151$.
We also set $\sigma = 0.8 \Delta x$.

To construct training data for eDMD, 
we collect snapshot pairs $\{(x^{(l)},x^{+(l)})\}_{l=1}^{M}$, with $x^{+(l)} = T_\mathrm{Duffing}(x^{(l)})$.
The initial samples $\{x^{(l)}\}_{l=1}^M$ 
are obtained by sampling $M = 100{,}000$ points
independently and uniformly from $[-2,2]\times[-1,1]$.

\subsubsection{Results}
We perform the state estimation for 
the discrete-time system $T_\mathrm{Duffing}$. 
We compare the performance of ePFOF with EnKF and PF, which are widely used 
methods for nonlinear state estimation.
The number of particles for PF and ensemble members for EnKF is varied over 
$25, 50, 100, 250, 500, 1000, 2000,$ and $4000$. 
Additionally, we apply a small process noise 
with the covariance of $1.0\times 10^{-4}I_2$ in the PF 
to prevent particle degeneracy.
We perform 200 independent state-estimation experiments 
with initial true states randomly selected from points
on the chaotic attractor of $T_\mathrm{Duffing}$. 
The initial distribution $\rho_0$ 
is set to the uniform distribution over $[-2,2]\times[-1,1]$.
All computations were performed in Python on a machine
equipped with an Intel Core i9-14900KF processor and
128 GB of RAM. For a fair comparison, the computational times
of all methods were measured on a single CPU core. 

Figure~\ref{fig:dist_duffing} illustrates how the distribution is updated during the 
first filtering iteration. 
The EnKF is a Monte Carlo-based approximation method.
Although the prior distribution is estimated from a large
ensemble, the analysis step is performed under the Gaussian
assumption. 
In contrast, the ePFOF employs a function-based
approximation, yielding both prior and posterior distributions
as smooth functions (specifically, a linear combination of
Gaussian RBFs) and explicitly accounts for
non-Gaussianity. 
The resulting posterior exhibits a curved
shape that naturally aligns with the chaotic attractor. 
Meanwhile, the PF relies on particle-based approximations for
both prediction and analysis. It successfully captures the
non-Gaussian posterior as a cloud of weighted particles,
analogous to the ePFOF.

Figure~\ref{fig:rmse_duffing} summarizes the
RMSEs of each setting of PF, EnKF and ePFOF and the corresponding 
computational time of the filtering step (prediction and analysis).
We calculate the $\mathrm{RMSE}_k$ over the interval $30 \leq k \leq 40$, 
where the estimated values have already converged to their steady-state values.
Across all settings, 
the smallest RMSE is achieved by PF with 4000 particles.
However, this setting requires a longer computational time than the other settings.
Although the RMSE of ePFOF is not substantially worse than that of PF, 
ePFOF significantly reduces the computational time of the filtering step.
In particular, when ePFOF is used with $N = 338$ basis functions, 
the RMSE increases by only about 20\% compared with PF using 4000 particles, 
whereas the computational time is reduced by more than 
99\% relative to that of PF.
It should be noted, however, that ePFOF involves a preliminary offline computation, as shown in Table~\ref{table:cput_duffing}.
The offline computational times include sampling of the training initial states, generation of training snapshot pairs, 
construction of the dictionary and Gram matrices, and computation of all quantities reused in the online filtering phase.
Compared with the online computational time shown in Fig.~\ref{fig:rmse_duffing},  the cost of this offline computation is not excessively large. 
Thus, we contend that ePFOF achieves a substantial reduction in computational time compared with PF. 

\begin{table}[tb]
\centering 
\caption{Offline computation time of ePFOF for the Duffing equation. \label{table:cput_duffing}}
\begin{tabular}{cc}
   \hline
Time (seconds)  &Basis size \\ 
\hline
   90 &338 \\ 
   99  &648 \\ 
   158  &1326 \\ 
   353  &2556 \\ 
   772  &5151 \\ 
   \hline
\end{tabular}
\end{table}

\subsection{Forced Swing Equation}
Our next target is the periodically forced swing equation as a model of a two-machine power grid \cite{ueda1998nonlinear}, given by
\begin{align}
  \left.
  \begin{aligned}
  \dot{x}_1&=x_3, \\ 
  \dot{x}_2&=x_4, \\ 
  \dot{x}_3 &= 0.2 
  - 0.4\sin(x_1 - x_2)\\ 
  &\quad - 1.4\sin(x_1 - \varepsilon\sin(\Omega t))
  - D_1 x_3\\ 
  \dot{x}_4 &= 0.8\left\{
  0.3
  - 0.4\sin(x_2 - x_1)\right.\\ 
  &\quad \left.- 1.4\sin(x_2 - \varepsilon\sin(\Omega t))
  - D_2 x_4
  \right\}
  \end{aligned}
  \right\},
  \label{eq:swing}
\end{align} 
with $D_1=D_2=0.005$, $\varepsilon=0.01$ and $\Omega=1.04$. 
The swing equation has the four-dimensional state $x=[x_1,x_2,x_3,x_4]^\top\in X:=[-\pi,\pi]_{\rm per}^2\times\mathbb{R}^2$ and exhibits transient dynamics that approach (at least) three different periodic attractors. 
Their basin structure is complicated as shown in Fig.~3 of \cite{ueda1998nonlinear}. 
Here, we aim to assess the performance of ePFOF \emph{during transients}, i.e., several steps from the initial time before reaching the attractors. 
This is also consistent with the model's motivation in power grids, that is, transient stability analysis \cite{chiang2011direct}. 
In the same manner as for the Duffing equation, the stroboscopic observation induces the discrete-time dynamical system described by a $\mathcal{C}^\infty$-diffeomorphism 
\begin{align*}
T_{\rm Swing}: X\to X.   
\end{align*}

We suppose that the observation directly measures only the states of the first machine with additive Gaussian noise:
$y_k = [x_{1k},x_{3k}]^\top + v_k, \quad v_k \sim \mathcal{N}_{0,0.16I_2}. $

\subsubsection{Set-Up}
The estimation of the PFO matrix $P$ requires $M$ snapshot pairs $\{(x^{(l)},x^{+(l)})\}_{l=1}^M$ with $x^{+(l)} = T_{\mathrm{Swing}}(x^{(l)})$ for $l=1,\ldots,M$. 
As mentioned above, the diffeomorphism $T_{\rm Swing}$ has at least three stable fixed points with basins of attraction. 
In particular, one of the fixed points corresponds to a stable periodic solution with the second kind\footnote{For instance, the time evolution of $x_1(t)$ is represented as $x_1(t)=\Omega t+\tilde{x}_1(t)$ with $\tilde{x}_1(t+2\pi/\Omega)=\tilde{x}_1(t)$.} that represents a stepping-out motion of synchronous machines. 
If the initial samples include a state that leads to the stable point, then the accuracy of $P$ in the eDMD framework may deteriorate. 
To reduce the degree of deterioration, by estimating the so-called stability region \cite{chiang2011direct} under $\varepsilon=0$, 
we set the training domain as 
$S := \left\{
x \in X \ \middle|\  V(x) < V_{\mathrm{c}}\right\},$
from which the initial samples are sampled. 
Here, $V(x)$ is called an energy function for the swing equation \eqref{eq:swing} under $\varepsilon=0$, given by 
\begin{align*}
V(x) &= \frac{1}{2}x_3^2 + \frac{1}{2\cdot0.8}x_4^2
- 0.2 x_1 - 0.3 x_2 \\ 
&\quad -0.4 \cos(x_1 - x_2) - 1.4\cos x_1 - 1.4\cos x_2 .
\end{align*}
The critical value $V_{\mathrm{c}}$ is defined as the second smallest value of 
$V$ among equilibrium points of \eqref{eq:swing} under $\varepsilon=0$. 
The initial samples $\{x^{(l)}\}_{l=1}^M$ are obtained by sampling $M = 166,443$ points 
independently and uniformly from the training domain $S$.

In this example, we consider two choices of the basis functions for eDMD, which are referred to as \emph{basis (i)} and \emph{basis (ii)}.

Basis (i) is the same as that 
used in the Duffing example: Gaussian RBFs whose centers are placed on a uniform grid over
$[-\pi,\pi]^2\times[-3,3]^2$ with grid spacing $\Delta x$. 
The grid points lying in the training domain $S$ are only retained. 
As in the Duffing example, we consider 
several choices of the grid spacing: 
$\Delta x = 0.62, 0.52, 0.44, 0.37, 0.31,$ and $0.26$, 
which correspond to $N = 610, 1253, 2430, 4869, 9879,$ and $19957$. 
The width parameter is set to $\sigma = 0.8\Delta x$.

\begin{figure}[tb]
\centering
\includegraphics[width=1.0\linewidth]{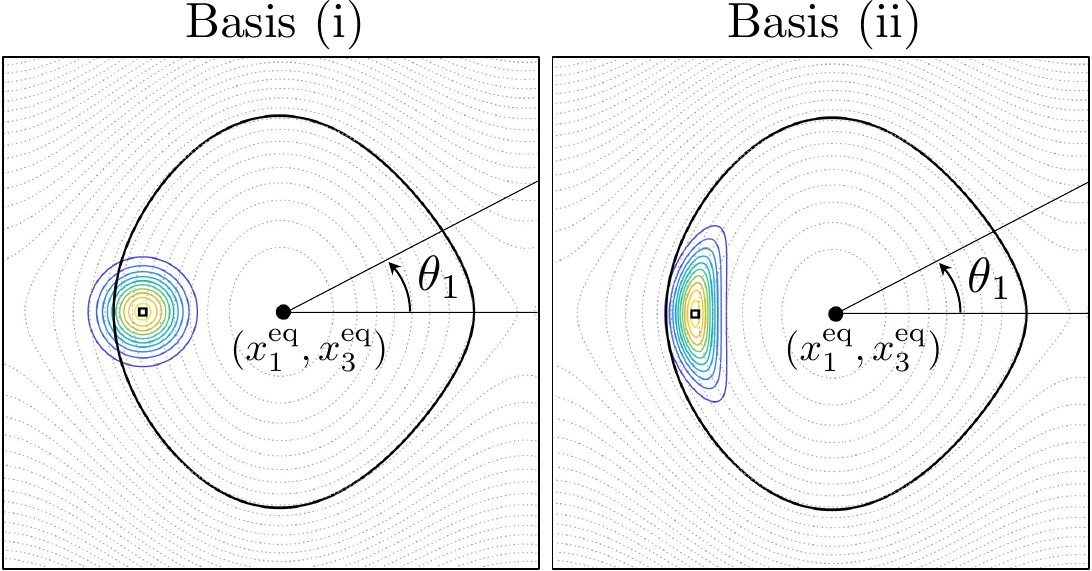}
\vspace{-8pt}
\caption{
   \label{fig:swing_basis}
   Examples of basis functions used in the forced swing equation.
   The left and right panels show basis (i) and basis (ii) on the $(x_1, x_3)$ plane.
   The colored contours indicate the values of the basis functions. 
   The center of the Gaussian RBFs is set to $(-1.5, 0,0,0)$.
   The black solid line stands for the contour of $V = V_c$,
   while the dotted lines show the contours of $V_1$. 
   The black dot represents the stable equilibrium point of \eqref{eq:swing} under $\varepsilon = 0$.
}
\end{figure}

Basis (ii) is synthesized to reflect the dynamical properties of the swing equation.
Although the swing equation contains small damping and forcing terms, 
if these effects are disregarded, 
that is, if $D_1 = D_2 = 0$ and $\varepsilon = 0$, 
the system can be described as a Hamiltonian system 
whose Hamiltonian is given by the energy function $V(x)$. 
Basis selection incorporating conserved quantities of Hamiltonian systems has been studied in Koopman analysis 
and modeling \cite{zhang2022hamiltonian,zhang2024PRR}, 
and has attracted attention as an effective approach 
for improving the accuracy and stability of numerical computations of the operators.
Motivated by this idea, we introduce basis functions 
that reflect the structure of the energy function as basis (ii). 
Specifically, we employ Gaussian RBFs defined with respect to the transformed coordinates:
\begin{align*}
   \psi_{\bar{x}_l,\sigma} = \exp \left(- \frac{\|w(x) - w{(\bar{x}_l)}\|^2}{2\sigma^2}\right), 
\end{align*}
where
\begin{align*}
   \left.
   \begin{aligned}
      w(x) &= [V_{1}(x), V_{2}(x), \theta_{1}(x), \theta_{2}(x)] \\ 
      V_{1}(x) &= \frac{1}{2}x_3^2 - 0.2 x_1 \\ 
      &\quad - \frac{0.4}{2} \cos(x_1 - x_2) - 1.4 \cos x_1 \\ 
      V_{2}(x) &= \frac{1}{2\cdot 0.8}x_4^2 - 0.3 x_2  \\ 
      &\quad - \frac{0.4}{2} \cos(x_2 - x_1) - 1.4 \cos x_2 \\ 
      \theta_1(x) &= \arg(x_1 - x_1^\mathrm{eq} + \mathrm{i}(x_3 - x_3^\mathrm{eq})) \\
      \theta_2(x) &= \arg(x_2 - x_2^\mathrm{eq} + \mathrm{i}(x_4 - x_4^\mathrm{eq})) 
   \end{aligned}
   \right\}.
\end{align*}
Here, 
$x^{\mathrm{eq}} = [x_1^\mathrm{eq}, x_2^\mathrm{eq}, x_3^\mathrm{eq}, x_4^\mathrm{eq}]^\top$ stands for the stable equilibrium point of \eqref{eq:swing} under $\varepsilon=0$. 
The functions $V_1(x)$ and $V_2(x)$ represent the energy components 
associated with the two machines in the power grid, 
and they satisfy $V_1(x) + V_2(x) = V(x)$. 
In the transformed coordinate system, the Gaussian RBFs are 
arranged along the level sets of the energy function
(see Fig. \ref{fig:swing_basis}).
Therefore, basis (ii) is
expected to represent the state distribution more appropriately than basis (i), 
which is defined by standard Gaussian RBFs in the original state coordinates.
The centers $\{\bar{x}_l\}_{l=1}^N$ are randomly selected from the subdomain
$\{x \in S \mid V(x) > -1.6\} \subset S$. 
The lower threshold on $V(x)$ is introduced to exclude centers near the stable equilibrium point $x^\mathrm{eq}$, where the Gaussian functions on the transformed coordinates become excessively narrow.
The width parameter is set to $\sigma = 0.3$. 
The number of RBFs is selected from 
$N = 500, 1000, 2000, 4000, 8000$, and $16000$.

\begin{figure*}[tb]
\centering
\includegraphics[width=0.8\linewidth]{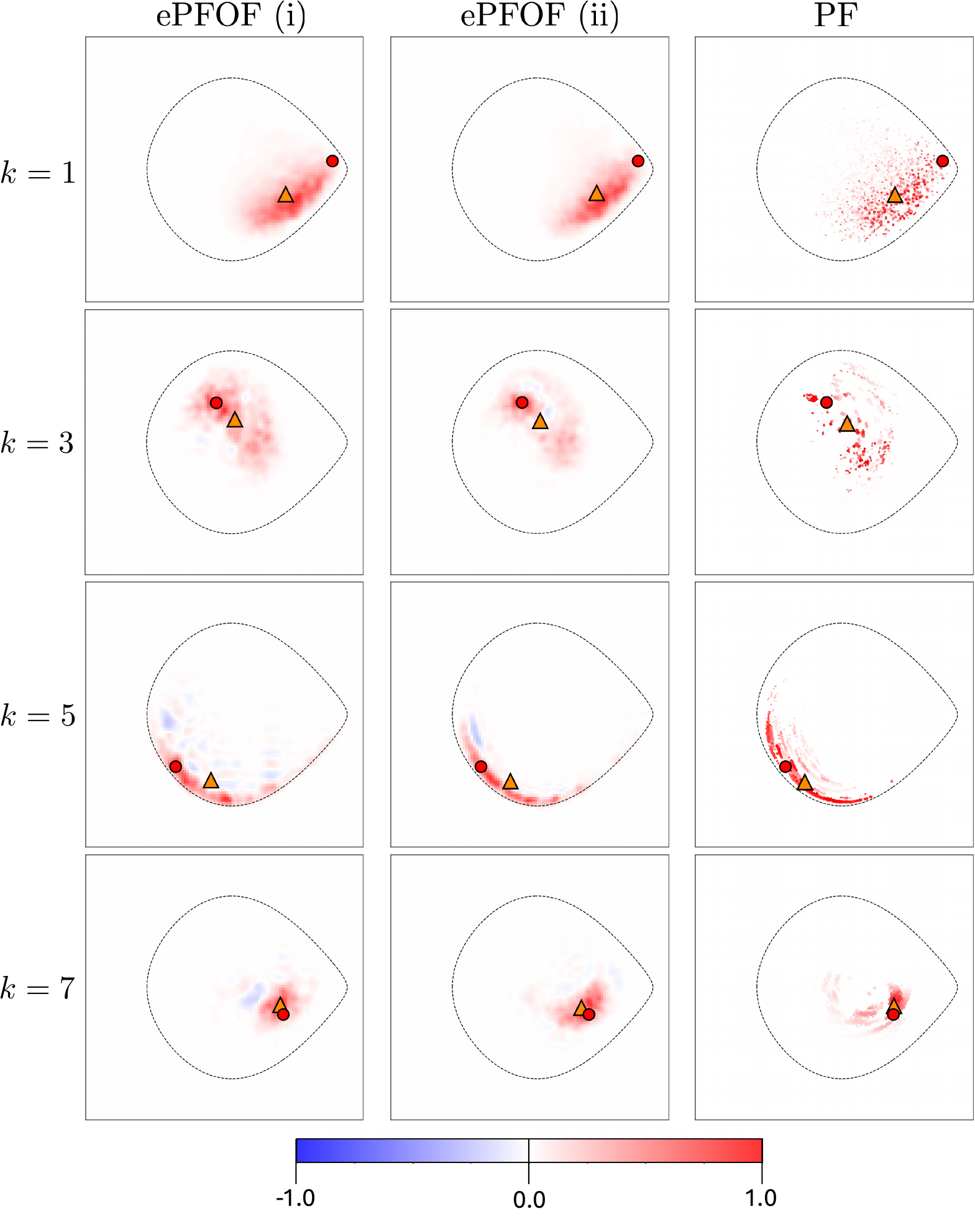}
\caption{
   Posterior distributions and state estimates for the forced swing equation. 
   From left to right, ePFOF (i), ePFOF (ii), and PF; from top to bottom, 
   the rows correspond to $k= 1,3,5,7$. 
   The distributions are marginalized over $x_1$ and $x_3$ 
   and plotted on the $(x_2, x_4)$ plane of the unobserved states.
   The black dash-dotted line shows the boundary of the
   training domain $S$ 
   and the overlaid colormap shows the posterior distributions
   normalized by their maximum absolute value, i.e., $\rho_{k|k}(x)/\max_{x \in X} |\rho_{k|k}(x)|$.
   The red point and orange
   triangle stand for the true state and MMSE
   estimate.
   \label{fig:dist_swing}
}
\end{figure*}

\begin{figure}[tb]
\centering
\includegraphics[width=1.0\linewidth]{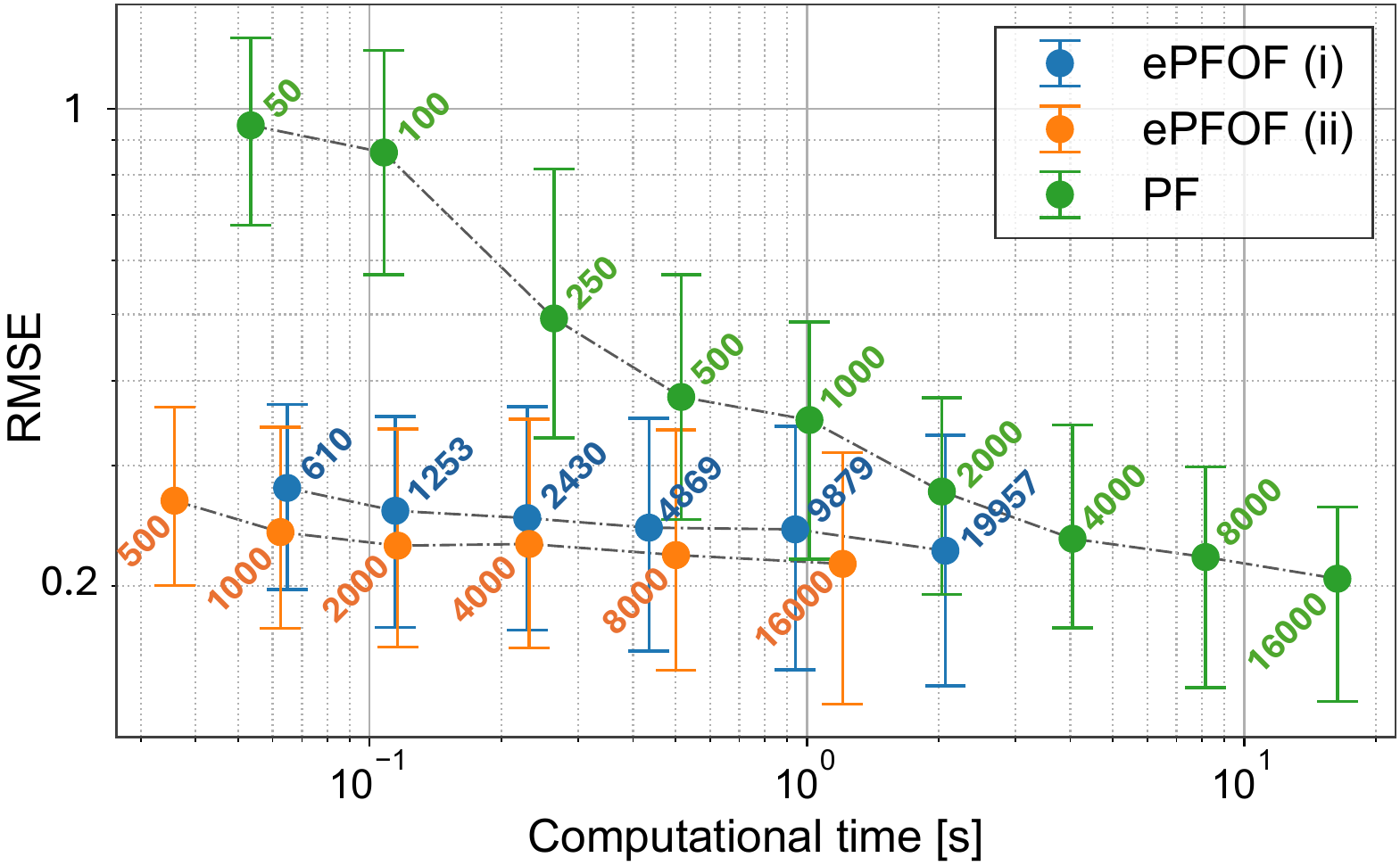}
\vspace{-10pt}
\caption{
   RMSE distributions at $k=7$ over 200 experiments, 
   plotted against the computational time per filtering step.
   Colored dots denote the medians, and error bars indicate the interquartile 
   ranges (25th-75th percentiles). The numbers next to the dots 
   indicate the number of particles for PF or 
   the number of basis functions for ePFOF.
   \label{fig:rmse_swing}
}
\end{figure}

\subsubsection{Results}
We perform state estimation for the discrete-time system $T_\mathrm{Swing}$. 
We compare the performance of ePFOF with PF, 
for which the number of particles 
is varied over $50, 100, 250, 500, 1000, 2000, 4000, 8000,$ and $16000$.
Although EnKF was considered in the previous example, it is not used as a benchmark here, since it failed to produce reliable estimates in this example due to divergence of the estimation results.
We conduct 200 independent state estimation experiments with randomly initialized true states. 
For each run, the initial true state is chosen to induce a transient non-Gaussian posterior. 
Specifically, the initial point is randomly selected from the set 
\begin{align*}
   \left\{x \in S \mid  \frac{V(x)- V(x^\mathrm{eq})}{V_\mathrm{c}- V(x^\mathrm{eq})} >0.95\right\}, 
\end{align*}
which corresponds to a high-energy region near the critical energy value $V_\mathrm{c}$. 
The initial distribution $\rho_0$ is set to the uniform distribution over the domain $S$.

Figure \ref{fig:dist_swing} illustrates 
the posterior distributions and 
the state estimates at time steps $k=1, 3, 5$, and $7$ for one of the 200 experiments,
where the numbers of basis functions are set to $19957$ for ePFOF with basis (i) (ePFOF (i)) and $16000$ for ePFOF with basis (ii) (ePFOF (ii)), 
and the number of particles in PF is set to $16000$. 
The overall shapes of the posterior distributions are broadly consistent among the three methods. 
As the filtering step proceeds, the posterior distribution gradually 
becomes more concentrated and elongated along the level set of the energy function, particularly at $k=5$ and $7$.
For ePFOF (i), the isotropic Gaussian basis functions cannot adequately represent such an elongated distribution. 
This leads to a large representation error, producing widely scattered negative regions, 
shown as blue regions in the panels for $k=5$ and $7$, which should not appear in a probability density function.
In contrast, ePFOF (ii) incorporates basis functions whose shapes follow the level sets of the energy function.
This enables ePFOF (ii) to capture the characteristic shape of the probability distribution more accurately. 
As a result, the representation error is reduced compared with ePFOF (i), and the negative regions are substantially suppressed.
These results suggest that basis design reflecting the properties of the target dynamical system 
is effective in reducing the representation error of density functions.
For PF, the posterior distribution is generally well represented by the particle ensemble. 
However, as shown in the panel at $k=3$, a reduction in particle diversity due to resampling is observed in some regions. 

Figure \ref{fig:rmse_swing} summarizes
the relationship between the RMSE and the computational time per filtering step (prediction and analysis)
for various settings of ePFOF and PF.
The RMSE distributions shown in the figure were obtained at $k=7$ from 200 independent experiments.
PF achieves good performance when a sufficiently large number of particles is used. 
However, its accuracy deteriorates substantially as the number of particles decreases.
In contrast, ePFOF maintains high accuracy even with a small number of basis functions, 
enabling accurate estimation within a short computational time. 
Moreover, ePFOF (ii) outperforms ePFOF (i), suggesting that the improved accuracy 
of the distribution computation shown in Fig.~\ref{fig:dist_swing} leads to a reduction in RMSE.
The corresponding offline computation times for ePFOF are listed in Table~\ref{table:cput_swing}. 
The offline computation can be completed within a few to a few tens of minutes, 
indicating that its computational cost is not prohibitive in this setting.

In practical power systems, 
measurements are obtained using sensing devices such as phasor measurement units (PMUs). 
However, because power systems are geographically distributed and subject to information-security constraints, 
it is generally difficult to directly observe all systems' states.
Therefore, online state estimation from limited observations is an important task 
for power-system monitoring and stable operation. 
The swing equation (14) is a fundamental model describing oscillatory dynamics and stability of power systems.
Despite its relatively simple formulation, 
it represents a strongly nonlinear system with multiple attractors 
and a complicated basin structure \cite{ueda1998nonlinear}. 
In the present numerical experiment, under a partial-observation setting 
in which only the states of the first machine are measured, 
the proposed method is able to estimate the states 
while representing a non-Gaussian state distribution. 
The results also show that the proposed method achieves estimation accuracy 
comparable to that of the benchmark particle filter while requiring a shorter online computation time.
These results suggest the potential applicability 
of the proposed method to state-estimation problems in strongly nonlinear systems, 
including power systems. 

\begin{table}[tb]
\centering 
\caption{Offline computation time of ePFOF for the swing equation. \label{table:cput_swing}}
\begin{tabular}{ccc}
   \hline
Method   &ePFOF (i)  &ePFOF (ii) \\ 
\hline
\multirow{6}{*}{\shortstack{Time (seconds) \\ / basis size}} 
   &179 / 610 &137 / 500\\ 
   &194 / 1253 &163 / 1000\\ 
   &232 / 2430 &218 / 2000\\ 
   &366 / 4869 &353 / 4000\\ 
   &870 / 9879 &753 / 8000\\ 
   &2901 / 19957 &1906 / 16000\\  
   \hline
\end{tabular}
\end{table}

\section{Conclusion}
\label{sec:conclusion}
We proposed the extended Perron--Frobenius Operator Filter (ePFOF) for nonlinear state estimation 
and validated its effectiveness through numerical simulations 
of the forced Duffing and swing equations. 
In regimes characterized by strong non-Gaussianity, 
ePFOF outperforms EnKF and PF.

As discussed in Section \ref{sec:ePFOF},
the finite-dimensional approximation obtained using eDMD
does not generally guarantee the non-negativity of
probability density functions.
The primary objective of the present study is
to evaluate the effectiveness of filtering
based on this finite-dimensional approximation
in terms of estimation accuracy and computational cost.
The numerical experiments demonstrate that
good estimation accuracy can be achieved
even when non-negativity was not guaranteed. 
This result indicates that
useful filtering performance can be achieved
without explicitly imposing positivity constraints.
Here, from a theoretical point of view, this does not imply that positivity constraints are unnecessary. 
We recognize the importance of exploring a novel formulation that allows us to control approximation error and computational cost while ensuring the mathematical rigor.
One possible direction is to build on the work of
Goswami et al. \cite{goswami2018constrained},
who proposed a constrained EDMD-based
approximation of the PFO using indicator functions
that preserves the stochastic-matrix structure.

The estimation performance of ePFOF strongly depends on the 
selection of basis functions. 
The results for the forced swing equation suggest that basis functions that  reflect the structure of the energy function of the target system
can help reduce errors in the estimated density functions and state estimates. 
This indicates that 
one important direction is to develop expressive basis functions that can reflect the intrinsic 
dynamical properties of the system.
This is closely related to basis learning methods that have been widely 
studied in the context of Koopman operator theory, including 
neural network-based learning~\cite{yeung2019learning,lusch2018}, analytically designed bases~\cite{netto2020}, and kernel-based methods~\cite{williams2015kernel}. 
Such approaches are also attractive from the viewpoint of efficient 
basis design for reducing the computational cost of ePFOF.
Lastly, its broader applications, including data assimilation for weather forecasts and control, are envisioned.

\appendix
We locate a coefficient vector $c_k^\ast\in\mathbb{R}^N$ such that the Galerkin conditions
\begin{align}
  \langle c_k^{\ast\top}\Psi, \psi_i\rangle_{\mathcal{L}^2} =  \langle \rho_k^\ast, \psi_i\rangle_{\mathcal{L}^2},
   \qquad i=1,\ldots,N, \label{eq:galerkin}
\end{align}
hold. 
Here, the inner product $\langle f,g\rangle_{\mathcal{L}^2}:= \int_X f(x)g(x)\,\mathrm{d}\mu(x)$ is used, 
and $\rho_k^*$ is the unnormalized posterior defined in \eqref{eq:rho-star}.
In practice, we approximate inner products by sample averages using data samples $\{x^{(l)}\}_{l=1}^M$ as 
\begin{align*}
   \displaystyle \langle f,g\rangle_{\mathcal{L}^2} \approx \frac{1}{M}\sum_{l=1}^M f(x^{(l)})g(x^{(l)}).
\end{align*}
Applying this empirical approximation, 
the conditions in \eqref{eq:galerkin} can be 
expressed as $Gc^*_k = G_\mathrm{W}c_{k|k-1}$, 
where $G$ is the Gram matrix defined in Section~\ref{sec:preliminaries}, 
and $G_\mathrm{W}$ is the observation-weighted Gram matrix given by
$\displaystyle
G_\mathrm{W} = \frac{1}{M}\Psi_X W_\ell \Psi_X^\top$, where $W_\ell = \operatorname{diag}(\ell_k(x^{(1)}), \ldots, \ell_k(x^{(M)})).
$
Applying  
$W_\ell$ to $\Psi^\top_X c_{k|k-1}$ 
is equivalent 
to the Hadamard product 
${e}_k \odot (\Psi^\top_X c_{k|k-1})$, 
where ${e}_k$ is the likelihood vector defined in \eqref{eq:p_k}. 
Thus, to avoid the explicit construction of $G_\mathrm{W}$, we simplify the right-hand side using this element-wise operation. 
Finally, for numerical stability and efficiency, 
rather than computing the inverse $G^{-1}$ directly, we employ the Cholesky factorization $G=LL^\top$ to solve the resulting linear equations, which directly yields \eqref{eq:c_ast}.

\section*{Funding}
This work was financialy supported by JST Moonshot R\&D Grant
Number JPMJMS2389 and JST BOOST Grant Number
JPMJBS2407.

\section*{Conflicts of Interest}
The authors declare no competing interests.

\section*{Author Contribution}
\begin{tabular}{ll}
   Conceptualization: &Y.~Miwa, Y.~Susuki \\ 
   Funding acquisition: &Y.~Miwa, Y.~Susuki, \\
   & S.~Kotsuki \\ 
   Investigation: &Y.~Miwa \\ 
   Supervision: &Y.~Susuki \\ 
   Visualization: &Y.~Miwa \\ 
   Writing--original draft: &Y.~Miwa, Y.~Susuki \\ 
   Writing--review \& editing: & Y.~Miwa, Y.~Susuki, \\ 
   &S.~Kotsuki
\end{tabular}

\end{document}